\newcommand{\be}{\begin{equation}}
\newcommand{\ee}{\end{equation}}

\documentclass{article}
\pdfoutput=1
\usepackage{amsmath}
\usepackage{amsfonts}
\usepackage{braket}
\usepackage{graphicx}
\usepackage[font={scriptsize,it}]{caption}
\usepackage{amssymb}
\usepackage{mathtools}

\begin{document}
\thispagestyle{plain}
\begin{center}
 \LARGE
A geodesic Witten diagram description of holographic entanglement entropy and its quantum corrections 
\vspace{0.7cm}
\large 

\textsc{Andrea Prudenziati \footnote{ andrea.prudenziati@gmail.com}}
\vspace{0.4cm}

\textsl{International Institute of Physics,\\ Universidade Federal do Rio Grande do Norte, \\Campus Universitario, Lagoa Nova, Natal-RN 59078-970, Brazil}
\vspace{0.9cm}
\end{center}


\begin{abstract}
We use the formalism of geodesic Witten diagrams to study the holographic realization of the conformal block expansion for entanglement entropy of two disjoint intervals. The agreement between the Ryu-Takayanagi formula and the identity block contribution has a dual realization as the product of bulk to boundary propagators. Quantum  bulk corrections instead arise from stripped higher order diagrams and back-reaction effects; these are also mapped to the structure for $G_N^0$ terms found in \cite{Faulkner:2013ana}, with the former identified as the bulk entanglement entropy across the Ryu-Takayanagi surfaces. An independent derivation of this last statement is provided by implementing a twist-line formalism in the bulk, and additional checks from the computation of mutual information and  single interval entanglement entropy. Finally an interesting correspondence is found between the recently proposed holographic entanglement of purification, and an approximated form for certain $1/c$ Renyi entropies  corrections.
\end{abstract}

\tableofcontents

\section{Introduction}
The celebrated Ryu-Takayanagi (RT) formula \cite{Ryu:2006bv} is a powerful tool that permits a holographic computation of entanglement entropy in conformal field theories (CFTs), at leading order in the gravity coupling constant $G_N$. An extensive amount of evidence has been collected so far, including various formal arguments for its validity (most notably \cite{Casini:2011kv} and \cite{Lewkowycz:2013nqa}). 
In particular one of the first tests that was devised covers the simplest case of a disconnected entangling region, the union of two disjoint intervals in a two dimensional CFT, \cite{Hartman:2013mia} and \cite{Headrick:2010zt}. Following \cite{Calabrese:2009qy} we can write the Renyi entropies $S^n(A\cup B)$ as a four points correlator of two twist and two anti-twist operators located at the boundary of the entangling region $\partial(A\cup B)=\partial A\cup \partial B$, and this can then be expanded in conformal blocks. Assuming exponentiation of the blocks in the large central charge limit $c\rightarrow \infty$ of CFTs with semi-classical holographic dual, the RT formula was matched at leading order in $1/c$ by the contribution of the identity block alone. 

A parallel dual construction was worked out by  \cite{Faulkner:2013yia}, starting from the representation of  Renyi entropies as a partition function over an $n$-sheeted replica trick manifold  $\Sigma=\mathbb{C}/\Gamma$, whose bulk counterpart is evaluated on a certain quotient geometry $AdS_3/\Gamma$, found following a procedure known as Schottky uniformization. In this language the classical RT result is the gravitational action on $AdS_3/\Gamma$ while one loop $1/c\approx G_N$ corrections are functional determinants, \cite{Barrella:2013wja}. 

The present paper pursues a slightly different path. We will still focus on the dual realization of the Headrick-Hartman idea but, differently from \cite{Faulkner:2013yia}, keeping the twist operator formalism and its conformal block expansion. The bulk construction will then make use of the so called geodesic Witten diagrams (gWds) \cite{Hijano:2015zsa}-\cite{Hijano:2015qja}, known to be the holographic duals of the conformal blocks.  Our goal is to study the properties of $1/c$ corrections to the classical RT formula in this formalism, which is particularly well suited to the task as it employs exclusively bulk objects. Explicitly these quantum bulk corrections will be expressed, at first order in $n-1$ \footnote{$n$ here and in the following is the replica index}, as a mixture of higher order gWds, corresponding to the propagation of light primary operators and their descendants, and back reaction effects on the geometry. 

This paper is organized as follows: in section \ref{background} we provide some background material and recap some useful formulas that will be used throughout the article. In section \ref{pertexpansion} we introduce the  expansion in conformal blocks of the Renyi entropies, and show that the RT result for the two intervals entanglement entropy is matched, at the CFT level, by the identity conformal block, and holographically by the four bulk to boundary propagators of the corresponding gWd, the rest of the diagram not contributing. In section \ref{purification} an interesting by-product result appears, the connection between certain bulk corrections to the Renyi entropies, computed as gWds of low enough scaling dimension (in a certain integrand approximation), and the holographic representation of the entanglement of purification, recently proposed by \cite{Takayanagi:2017knl}. In section \ref{fftt} we start our analysis on sub leading $1/c$ quantum bulk terms, first by considering the limit when the two intervals entanglement entropy, that indeed posses such corrections, reduces to the exact single interval case. These are then studied in detail in section \ref{interpret} where the general form at order $O(n-1)$ is  given and compared to the findings of \cite{Faulkner:2013ana}. There the main result is perhaps the identification of the bulk entanglement entropy across the RT surfaces as a sum of gWds stripped of the bulk to boundary propagators. In particular a twist line formalism is introduced to explain this result from a purely bulk perspective. We also provide a first test by computing the bulk entanglement entropy for a single interval RT surface, verifying that it vanishes and showing consistency among the different representations used (gravity replica trick determinant, twist line representation or gWds expansion). Furthermore contributions to the entanglement entropy from the first sub leading gWd are carefully worked out from previous CFT results as \cite{Calabrese:2009ez}-\cite{Calabrese:2010he}, and used in a second test to provide a match for the expected behaviour of mutual information. We conclude with a discussion on future directions for further research.

\section{Some background}\label{background}
We use as a reference \cite{Hijano:2015zsa}-\cite{Hijano:2015qja} \footnote{For simplicity, and because the two dimensional factorization into holomorphic and antiholomorphic part of the conformal blocks will not play any major role in this discussion, we will keep the notation used for generic dimensionality}. A four point amplitude in  CFT can be written as:
\be \label{uno}
\braket{O_1(x_1)O_2(x_2)O_3(x_3)O_4(x_4)}=\sum_{\Delta,l}C_{\Delta,l \; 12}C^{\Delta,l}_{34} W_{\Delta,l}(x_i),
\ee
with $\Delta$ and $l$ the scaling dimension and spin of the propagating primary operators. Moreover conformal invariance restricts the partial waves to have this form ($x_{ij}=x_i-x_j$)
\be \label{cinque}
W_{\Delta,l}(x_i)=\left(\frac{x_{24}^2}{x_{14}^2} \right)^{\frac{1}{2}\Delta_{12}}\left(\frac{x_{14}^2}{x_{13}^2} \right)^{\frac{1}{2}\Delta_{34}}\left(x_{12}^2\right)^{-\frac{1}{2}(\Delta_1+\Delta_2)}\left(x_{34}^2\right)^{-\frac{1}{2}(\Delta_3+\Delta_4)}  G_{\Delta,l}(u(x_i),v(x_i)),
\ee
being $u,v$ the cross-ratios of the distances $x_{ij}$:
 \[u=\frac{x_{12}^2 x_{34}^2}{x_{13}^2 x_{24}^2}, \;\;\; v=\frac{x_{14}^2 x_{23}^2}{x_{13}^2 x_{24}^2}. \]
 Defining for $l=0$:
\be \label{quattro}
W_{\Delta,0}(x_i) =W_{\Delta}(x_i)=\mathcal{W}_{\Delta}(x_i)\frac{2\Gamma(\Delta)}{\Gamma(\frac{\Delta+\Delta_{12}}{2})\Gamma(\frac{\Delta-\Delta_{12}}{2})}\frac{2\Gamma(\Delta)}{\Gamma(\frac{\Delta+\Delta_{34}}{2})\Gamma(\frac{\Delta-\Delta_{34}}{2})}
\ee
we can compute holographically these objects using a geodesic Witten diagram \footnote{this can also be generalized to non zero spin by replacing the bulk to bulk propagator $G_{bb}^{\Delta}$ with its more complicated spin-$l$ cousin, pulled back on the two geodesics. See \cite{Hijano:2015zsa} for details. }
\be \label{due}
\mathcal{W}_{\Delta}(x_i)=\int_{\gamma_{12}}\hspace{-0.2cm}d\lambda\int_{\gamma_{34}}\hspace{-0.2cm}d\lambda' G_{b\partial}(y(\lambda),x_1)G_{b\partial}(y(\lambda),x_2)G_{bb}^{\Delta}(y(\lambda),y'(\lambda'))G_{b\partial}(y'(\lambda'),x_3)G_{b\partial}(y'(\lambda'),x_4).
\ee
 Notation is as follows: $\gamma_{ij}$ are geodesics starting and ending at the boundary points $x_i,x_j$, parametrized by proper time $\lambda,\lambda'$. The four bulk to boundary propagators $G_{b\partial}$ are divided into two couples corresponding to the OPE channel of (\ref{uno}). Within each couple the two $G_{b\partial}$ start at $x_i$ and $x_j$ respectively and end on a common bulk point $y(\lambda)$ (or $y'(\lambda')$), along $\gamma_{ij}$. The bulk to bulk propagator $G_{bb}$ connects $y$ with $y'$. Note that the above formula evaluated in AdS is valid in the limit of light boundary operators, without including back reaction on the metric. 

In Euclidean Poincar\'e coordinates the boundary to bulk propagator is
\be \label{sette}
G_{b\partial}(y(u,x),x_i)=\left(\frac{u}{u^2+|x-x_i|^2} \right)^{\Delta_i}
\ee
and the bulk to bulk one is
\be \label{dieci}
G_{bb}^{\Delta}(y,y')=e^{-\Delta \sigma(y,y')} \prescript{}{2}F_1\left(\Delta,\frac{d}{2};\Delta+1-\frac{d}{2};e^{-2 \sigma(y,y')} \right),
\ee
with $\sigma(y,y')$ the geodesic distance between $y$ and $y'$.

We will consider two dimensional CFTs and choose a very specific form for the correlator (\ref{uno}), containing two couples of twist and anti-twist operators \footnote{for definitions of twist operators and they relation with the replica trick procedure and Renyi entropies we refer the reader to \cite{Calabrese:2009qy}, here we assume general knowledge of these ideas.}. This four point amplitude is used to compute the Renyi entropies for a disconnected entangling region $A\cup B$ with $\partial(A\cup B)=\{x_1,x_2,x_3,x_4\}$ \footnote{\label{foot}the correlator $\braket{\tau_n(x_1)\tau_n^{-1}(x_2)\tau_n(x_3)\tau_n^{-1}(x_4)}$ is computed as a path integral on the replica manifold with the original Lagrangian $Z_n$, normalized by a path integral on $n$ non-interacting copies of the original manifold, $Z_1^n$: \[ \braket{\tau_n(x_1)\tau_n^{-1}(x_2)\tau_n(x_3)\tau_n^{-1}(x_4)}=\frac{Z_n}{Z_1^n}\propto Tr_{(AB)^c}\rho_{AB}^n. \] The proportionality constant is usually indicated as $c_n$, such that $c_1=1$, and an UV cutoff can also appear.   }:
\be \label{reppath}
\braket{\tau_n(x_1)\tau_n^{-1}(x_2)\tau_n(x_3)\tau_n^{-1}(x_4)}\propto Tr_{(AB)^c}\rho_{AB}^n.
\ee
The scaling dimension is the same for both twist and anti-twist operators \footnote{\label{foot2} conventions  as follows: the conformal dimension $h,\bar{h}$ of two dimensional CFT operators is related to the scaling dimension $\Delta$ and spin $l$ as $\Delta=h+\bar{h}$ and $l=h-\bar{h}$.} 
\be\label{tre}
\Delta_{\tau_n}=\Delta_{\tau^{-1}_n}=\Delta_n=\frac{c}{12}\left(n-\frac{1}{n}\right).
\ee
These operators are not light for generic $n$ as the semi-classical holographic limit sends the central charge to infinity, $c \rightarrow \infty$. However if we first pick the appropriate limit $n\rightarrow 1$ (see section \ref{nc}) the dimension above can be kept small. 

The entanglement entropy can be recovered as:
\be \label{ee2}
S(AB)=\partial_n S^n(AB)|_{n=1}
\ee
with
\be \label{ren2}
S^n(AB)=-Tr_{(AB)^c}\rho_{AB}^n.
\ee
The holographic RT formula for (\ref{ee2}) predicts that, at classical level in the bulk, the entanglement entropy $S(AB)$ is proportional to the sum of the two RT surfaces that, in this case, are the geodesics connecting the points on $\partial(A\cup B)$ and such that their total length is the minimal one. To simplify formulas we will pick the four points $x_1, x_2, x_3, x_4$ to be real and ordered as $x_1<x_2<x_3<x_4$, with twist operators at $x_1$ and $x_3$ and anti twist operators at $x_2$ and $x_4$. With these choices either $\partial A=x_1\cup x_2$ and $\partial B=x_3\cup x_4$ or $\partial A=x_1\cup x_4$ and $\partial B=x_2\cup x_3$ but, assuming  $\text{length} (\gamma_{12})+\text{length}(\gamma_{34})<\text{length} (\gamma_{14})+\text{length}(\gamma_{32})$, we always have:
\be \label{rtee2}
S(AB)=\frac{\text{length}(\gamma_{12})+\text{length}(\gamma_{34})}{4 G_N}=\frac{c}{3}(\log |\frac{x_{12}}{\epsilon}|+\log |\frac{x_{34}}{\epsilon}|).
\ee
Here  $\epsilon$ is a UV cutoff and the relation between the central charge and the Newton constant in three dimensional gravity in AdS with radius $L$ has been used: $c=\frac{3 L}{2 G_N}$.

Before moving on let me discuss the choice of contraction in (\ref{uno}). In general this choice should be irrelevant, and the full result independent; in our case however, as we will concentrate only on the first few terms, it is convenient to fix the convention that the OPE contraction will always be done accordingly to the RT formula, that is twist anti-twist contraction as $1\leftrightarrow 2$ and $3\leftrightarrow 4$ (instead that $1\leftrightarrow 4$ and $2\leftrightarrow 3$). In this way we will be able to match perturbatively the RT result, that otherwise would be recovered only non perturbatively in $1/c$ in the "wrong" channel. 

\section{Conformal block expansion and geodesic Witten diagrams}\label{pertexpansion}
We start by using the expression for the  Renyi entropies (\ref{ren2}) and its relationship with (\ref{reppath}). The four points correlator can be expanded in conformal blocks of increasing scaling dimension (and later on spin) and these computed either from CFT formulas or holographically, by using the gWd formulation (\ref{due}). Generically speaking this second option is more interesting if the goal is to understand how the RT formula and its leading corrections arise, as it already lives in the bulk.

As in \cite{Hartman:2013mia} and \cite{Headrick:2010zt} we assume that the CFT should posses an holographic dual theory in the semi classical $c\rightarrow \infty$ limit, and expected related properties: for example keeping the  number of primary operators below a certain fixed dimension finite in this limit, and the growth  of the OPE coefficients in the central charge less then exponential, to not offset the conformal blocks (more general discussions can be found  in either  \cite{Hartman:2013mia},  \cite{Hartman:2014oaa} or \cite{Headrick:2010zt}). We then have that the identity block provides the leading $c$ contribution to the RT formula while $1/c$ corrections come from the remaining low dimension conformal block terms \footnote{ here I should also mention the results of \cite{Belin:2017nze} where it was shown that the growth of the OPE coefficients for multi-trace operators is exponential, which may mean that the vacuum block contribution is no longer dominant and a large number of other blocks should be taken into account. Yet it is not clear if this effect is relevant for Renyi entropies with $n\simeq 1$. }. This will be our starting point. The goal then is to transform these CFT statements into a purely bulk relation between gWds for different primary operators (plus back-reaction on the geometry) and the holographic description of entanglement entropy. 

With these premises we use (\ref{ren2}) to write
\be \label{alt}
S(AB)=\partial_n S^n(AB)=\sum_{\Delta,l}\partial_n S ^n(AB))|_{\Delta,l}
\ee
with
\be \label{alt2}
S^n(AB))|_{\Delta,l}=-C_{\Delta,l \; 12}C^{\Delta,l}_{34} W_{\Delta,l}.
\ee

\subsection{A note on twist operators and holography}
Before moving on I would like to clarify what we mean by the holographic dual of twist operators and their OPE. 

In the twist operator formalism the $n$-sheets of the replica trick construction collapse into a single one hosting the $n$-decoupled identical Lagrangians that were living on the $n$ sheets. However the fields do maintain a local coupling at the twist operator insertions.

The OPE for twist operators takes the general form \cite{Calabrese:2009ez}-\cite{Calabrese:2010he}
\be \label{genope}
\tau_n(x)\tau_n^{-1}(y)=\sum_{\{k_j\}}C_{\{k_j\}}\prod_{j=1}^n O_{k_j}(z)
\ee
where $O_{k_j}$ are a complete set of operators for the $j$th theory, $j=1,\dots,n$.

Because of the operator mixing the conformal block expansion of (\ref{reppath}) does not in general factorize in the index $j$, nor in the OPE coefficients and neither in the conformal blocks. Still there should be some kind of j-factorization away from the twist operator insertions, in the propagating operators. 

The holographic dual of $n$ decoupled identical CFTs in the vacuum (without entanglement between different theories) is given by the corresponding $n$  decoupled bulk gravity Lagrangians on a common AdS background.  But the twist operators belong to a different theory which is the replicated theory constructed on a periodically identified multi-sheet space. The recipe for finding the holographic dual of this boundary replica trick construction  was provided by \cite{Lewkowycz:2013nqa}. There they considered geometries with compactified euclidean time along a circle that, at the boundary, would surround the end points of the entangling regions, that is the twist operator insertions; in the bulk this circle collapses at the RT surfaces. In this way the boundary location of the twist operators is extended in the bulk, which will be a key ingredient in identifying, later on in section  \ref{flmquantum}, the bulk dual of a couple of twist and anti-twist operators as a bulk twist line on the corresponding RT surface. If this identification is accepted then the CFT exchange of operators along the conformal block channel, following the OPE (\ref{genope}), transforms into a bulk exchange of the fields dual to the ones appearing in (\ref{genope}), from one twist line at one RT surface to the other. This is the exact form that the Geodesic Witten diagrams have, provided that the bulk operator of dimension $\Delta$ entering the bulk to bulk propagator (\ref{dieci}) is really understood as a collection of the bulk fields $\phi_{k_j}$, dual to the boundary operators $O_{k_j}$. Indeed writing $\Delta=\sum_j \Delta_k$ we can immediately check that the propagator (\ref{integ}) factorizes as
\be\label{factorprop}
G_{bb}^{\Delta}(y,y')=G_{bb}^{\Delta_1}(y,y')\dots G_{bb}^{\Delta_n}(y,y').
\ee
This bulk propagator factorization in the $j$-index is exactly what we meant as factorization away from the twist operator positions, at the beginning of this section.

As we will not actually need to use this factorized version of the propagator we will instead keep using in the following the simple notation $G_{bb}^{\Delta}(y,y')$ to denote the right hand side of (\ref{factorprop}). These ideas will be developed in more detail in section \ref{interpret}.

\subsection{Zeroth order is the Ryu-Takayanagi result: CFT computation}\label{zeroth}
We start with a purely CFT computation and consider the lowest scaling dimension among primary operators in the OPE expansion of $\tau$ and $\tau^{-1}$: the identity for which $\Delta=\Delta_0=0$. The general result of \cite{Hartman:2013mia}  is that, by assuming exponentiation of the identity block in the large central charge limit, the RT result for the two interval entanglement entropy is matched at leading order in $c$ and at finite value of the distance between the intervals \footnote{ up to the phase transition point where you need either to know all the non-perturbative terms or change the expansion channel as we do here. }. In the present section we will not assume any specific behaviour for the identity block nor take any explicit $c\rightarrow \infty$ limit; in this case obviously the control over $1/c$ corrections is lost, so the result is not particularly meaningful. In the next section however, by considering a gWd formalism, holographic properties are automatically taken into account, and previously described results on $1/c$ terms valid, as long as the gWds accurately describe the corresponding conformal blocks.

With this in mind let us consider the known CFT expression for $G_{\Delta}(u,v)$ as $\Delta\rightarrow 0$, for the case of four twist operators when $n\rightarrow 1$.  Here and in the following $l=0$ for most of the time, so this index will remain hidden until further notice. The entanglement entropy at this order is
\be \label{sei2}
S(AB)|_{0}=-\partial_n \left(C_{\Delta_0 \; 12}C^{\Delta_0}_{34} W_{\Delta_0} \right)|_{n=1},
\ee
that using (\ref{cinque})
\be
W_{\Delta_0}=\left(|x_{12}|^2\right)^{-\Delta_n}\left(|x_{34}|^2\right)^{-\Delta_n}  G_{\Delta_0}(u,v)
\ee
becomes
\be 
S(AB)|_0=-\partial_n \left[ C_{\Delta_0 \; 12} C^{\Delta_0}_{34} G_{\Delta_0}(u,v) |x_{12}|^{-2\Delta_n}|x_{34}|^{-2\Delta_n} \right]_{n=1}.
\ee
In particular the terms containing explicit dependence on $n$ are:
\be
-\partial_n \left[ |x_{12}|^{-2\Delta_n}|x_{34}|^{-2\Delta_n} \right]_{n=1}=\frac{c}{6}(\log |x_{12}|+\log |x_{34}|)\partial_n\left(n-\frac{1}{n}\right)_{n=1}\hspace{-0.3cm}=\frac{c}{3}(\log |x_{12}|+\log |x_{34}|).
\ee
This result reproduce  the Ryu-Takayanagi formula  for the two intervals entanglement entropy (\ref{rtee2}) \footnote{apart from the UV cutoff to which the replica trick procedure is essentially blind and should be added by hand.}, provided that 
\be \label{con1}
C_{\Delta_0 \; 12}|_{n=1}= C^{\Delta_0}_{34}|_{n=1}= G_{\Delta_0}(u,v)|_{n=1}= 1
\ee
and
\be\label{con2}
\partial_n C_{\Delta_0 \; 12}|_{n=1}= \partial_n C^{\Delta_0}_{34}|_{n=1}= \partial_n G_{\Delta_0}(u,v)|_{n=1}=0.
\ee
The OPE coefficients  $C_{\Delta_0 \; 12}$, $C^{\Delta_0}_{34}$ are easily evaluated considering that, by definition, when the twist and anti-twist operators collide, they annihilate each other with no other effect: $\lim_{r\to 0}\tau_n(r)\tau_n^{-1}(0)= Id $ for every $n$, so that both (\ref{con1}) and (\ref{con2}) are satisfied (here the OPE coefficients do not depend on the distance, which is included in $W_{\Delta,l}(x_i)$). We want then to compute $G_{\Delta_0}(u,v)$ and $ \partial_n G_{\Delta_0}(u,v)|_{n=1}$. One simple way to do this is by using the CFT formula for its integral representation in $d$ dimensions and generic scaling $\Delta$:
\be \label{conf0}
G_{\Delta}(u,v)=\frac{\Gamma(\Delta)}{\Gamma(\frac{\Delta+\Delta_{34}}{2})\Gamma(\frac{\Delta-\Delta_{34}}{2})}u^{\Delta /2}\int_0^1 d\sigma \sigma^{\frac{\Delta+\Delta_{34}-2}{2}}(1-\sigma)^{\frac{\Delta-\Delta_{34}-2}{2}}(1-(1-v)\sigma)^{\frac{-\Delta+\Delta_{12}}{2}}\cdot
\ee 
\[
\cdot \prescript{}{2}F_1\left(\frac{\Delta+\Delta_{12}}{2},\frac{\Delta-\Delta_{12}}{2};\Delta-\frac{d-2}{2};\frac{u\sigma(1-\sigma)}{1-(1-v)\sigma} \right).
\]
This expression simplifies considerably at  $d=2$ and with the four scalar boundary operators having the same dimension. The only dependence on $n$  from the scaling $\Delta_n$ then drops out from the above formula, so we can immediately fulfil (\ref{con2}). Moreover using the series definition for the Hypergeometric function we can easily show that $\prescript{}{2}F_1\left(\frac{\Delta_0}{2},\frac{\Delta_0}{2};\Delta_0;z \right)\xrightarrow[\text{$\Delta_0\rightarrow 0$}]{} 1$.
Then (\ref{conf0}) becomes (only the relevant terms of $\Delta_0$ are retained in the limit)
\be \label{intrep}
G_{\Delta_0}(u,v)=\lim_{\Delta_0 \to 0}\frac{\Gamma(\Delta_0)}{\Gamma(\frac{\Delta_0}{2})^2}\int_0^1 d\sigma \sigma^{-1+\frac{\Delta_0}{2}}(1-\sigma)^{-1+\frac{\Delta_0}{2}}.
\ee 
Given the integral representation
\be \label{undici}
\frac{\Gamma(x)\Gamma(y)}{\Gamma(x+y)}=\int_0^1 d\sigma \sigma^{x-1}(1-\sigma)^{y-1}
\ee
also the result (\ref{con1}) follows immediately for $G_{\Delta_0}(u,v)$. So we have proven that
\be \label{dodici}
S(AB)|_0=\frac{c}{3}(\log |x_{12}|+\log |x_{34}|).
\ee

\subsection{Holographic computation}\label{nc}
We want here to show that the contribution from the gWd for the identity conformal block produces the RT result as in the previous section. Once more let me stress that, in doing so, we automatically restrict to CFTs with holographic properties, and assume the validity of the gWd to reproduce the conformal block expression at any order in $1/c$. Then previous results for holographic CFTs connecting the identity block to the RT formula, and low scaling blocks to $1/c$ corrections, apply. Moreover, working directly in the bulk, we will no longer just compare formulas in between the two sides of the AdS/CFT duality, but directly pick the objects that correspond to the RT curves, and later on its corrections.

Here it is important to discuss a little bit more the effect of placing twist and anti-twist operators at the boundary. As we are now considering an holographic theory, the large central charge limit $c\rightarrow \infty$ should be taken, so for generic $n$ the twist operator dimension (\ref{tre}) is large, and these do indeed back-react on the geometry. Fortunately in the vicinity of $n=1$ the dimension $\Delta_n$ can be taken to be as small as required, even for large central charge \footnote{For example we can parametrize the real $n$ as $n\rightarrow m/c+1$, and then achieve the limit $n\rightarrow 1$ as $m\rightarrow 0$ and/or $c\rightarrow \infty$ (obviously for finite $n$ we would need $m$ to growth faster then $c$). This parametrization is interesting because the two limits coincide when computing the scaling dimension:
\begin{align}
\Delta_n=\frac{c}{12} \left(n-\frac{1}{n}\right) =\frac{1}{12}\frac{m(m+2c)}{m+c}\xrightarrow[\text{$m\rightarrow 0$}]{}\frac{m}{6} \\
\xrightarrow[\text{$c \rightarrow \infty$}]{}\frac{m}{6} \nonumber .
\end{align} }. The problem is that, to obtain the analytic continuation in $n$ required for computing the Renyi entropies nearby $n=1$, we need in general the results for all integer $n$ (plus other inputs, see for example the discussion in \cite{Calabrese:2010he}). As for these values the twist operator scaling \emph{is} large and there is back reaction, it may appear we are in a loop hole argument. The strategy will then be as follow: we first consider the holographic description of the four twist operators correlator  at first order in $(n-1)$. This is a well posed problem at this level as the analytic continuation of the scaling dimension $\Delta_n$ is trivial. At this order we can broke down the  holographic Renyi entropy in various pieces, and separately consider each of them. 
We have two contributions: the first piece comes from direct application of (\ref{quattro}) and (\ref{due}) and it is of order $(n-1)^0$ in the geometry (no back reaction considered, the geometry being just AdS) but the OPE and the holographic evaluation of conformal blocks will be done at order $(n-1)$. The second contribution is from the holographic description of conformal blocks in the back reacted geometry that we can call $M_n$ (so $M_0=\text{AdS}$); here the geometry is at order $(n-1)$ (for example in the heavy-light limit the deficit angle would be taken small at order $(n-1)$) but the OPE is evaluated at order $(n-1)^0$, so only the identity propagates. Finally we also have the $(n-1)^0$ term which is a gWd in AdS with propagation of the identity alone. Everything else is higher order in $(n-1)$ and is not important for the goal of computing EE. In summary:
\be \label{expn1}
\braket{\tau_n(x_1)\tau_n^{-1}(x_2)\tau_n(x_3)\tau_n^{-1}(x_4)}|_{\text{up to}\; O(n-1)}=[\text{AdS},\Delta=0 ](=\text{RT})+
\ee
\[
+[\text{AdS},\Delta\approx O(n-1)](=\text{gWd}_{O(n-1)})+[M_n\approx O(n-1),\Delta=0](=\text{Id} \;\text{in}\;\;\text{backreacted}_{O(n-1)}).
\]
This subdivision in three terms is valid independently of the explicit form of analytic continuation of the Renyi entropies, it is just a first order expansion around $n=1$. The idea now is to consider the  analytic continuation for each piece separately around $n=1$, which is a more complicated story.  For the moment let us concentrate on the $(n-1)^0$ term that was the subject of the previous section.

Knowing that  $C_{\Delta_0 \; 12}= C^{\Delta_0}_{34}=1$ and using (\ref{quattro}) we obtain
\be
S(AB)|_{0}=-\partial_n \left(\mathcal{W}_{\Delta_0}\frac{4\Gamma(\Delta_0)^2}{\Gamma(\frac{\Delta_0}{2})^4}\right)_{n=1}.
\ee

The evaluation of $\mathcal{W}_{\Delta_0}$ will now be done holographically by applying (\ref{due}). We start with the bulk to boundary propagators $G_{b\partial}(y,x_i)$ of (\ref{sette}). Working with Poincar\'e coordinates the geodesics $\gamma_{ij}$ are described by the equation
\be 
2u^2 +(x-x_j)(\bar{x}-\bar{x}_i)+(x-x_i)(\bar{x}-\bar{x}_j)=0
\ee
that is solved by
\begin{align} \label{otto}
x(\lambda)=& \frac{x_i+x_j}{2}+ \frac{x_i-x_j}{2}\tanh \lambda \nonumber \\
\bar{x}(\lambda)=& \frac{\bar{x}_i+\bar{x}_j}{2}+ \frac{\bar{x}_i-\bar{x}_j}{2}\tanh \lambda  \\
u(\lambda)=& \frac{|x_i-x_j|}{2\cosh \lambda}\nonumber .
\end{align}
Inserting these into (\ref{sette}) we obtain the simple answer:
\be \label{bp}
G_{b\partial}(y(\lambda),x_1)G_{b\partial}(y(\lambda),x_2)=\left(|x_1-x_2|^2\right)^{-\Delta_n}
\ee
and likewise for $(1,2)\leftrightarrow (3,4)$. So the four bulk to boundary propagators contribute to a common factor in the integrand \emph{independent} of the actual bulk points along the geodesics. This result is not affected by the choice of $\Delta$ in the intermediate channel, and so will be valid for all the terms in (\ref{alt}). The bulk to bulk propagator instead does obviously depend on $\Delta$ and its form is given by (\ref{dieci}). In our case it is evaluated as: 
\be \label{integ}
G_{bb}^{\Delta}(y,y')= e^{-\Delta \sigma(y,y')}\prescript{}{2}F_1\left(\Delta,1;\Delta;e^{-2 \sigma(y,y')} \right)=
\ee 
\[ 
=\frac{e^{-\Delta \sigma(y,y')}}{1-e^{-2 \sigma(y,y')}}\prescript{}{2}F_1\left(0,1;\Delta;\frac{e^{-2 \sigma(y,y')}}{e^{-2 \sigma(y,y')}-1} \right)=\frac{e^{-\Delta \sigma(y,y')}}{1-e^{-2 \sigma(y,y')}}
\]
where we used a Pfaff transformation to pass to the second line.
The result for $\mathcal{W}_{\Delta_0}$ than is
\be \label{perso}
\mathcal{W}_{\Delta_0}= \left(|x_1-x_2|^2\right)^{-\Delta_n}\left(|x_3-x_4|^2\right)^{-\Delta_n}\int_{\gamma_{12}}\hspace{-0.3cm}d\lambda\int_{\gamma_{34}}\hspace{-0.3cm}d\lambda' \frac{e^{-\Delta_0 \sigma(y(\lambda),y'(\lambda'))}}{1-e^{-2 \sigma(y(\lambda),y'(\lambda'))}}
\ee
that agrees with our previous result (\ref{dodici}), after deriving in $n$ and plugging in the missing factors of Gamma functions of (\ref{quattro}), provided that
\be \label{exp}
\lim_{\Delta_0\to 0}4\frac{\Gamma(\Delta_0)^2}{\Gamma(\frac{\Delta_0}{2})^4}\int_{\gamma_{12}}\hspace{-0.3cm}d\lambda\int_{\gamma_{34}}\hspace{-0.3cm}d\lambda' \frac{e^{-\Delta_0 \sigma(y(\lambda),y'(\lambda'))}}{1-e^{-2 \sigma(y(\lambda),y'(\lambda'))}}= 1.
\ee

In practice the RT formula is reproduced by the contribution from the four bulk to boundary propagators alone! So we see here a refinement of the result that the identity block reproduces the RT formula, being able to isolate precisely the bulk objects responsible for this.
 
Let us now prove (\ref{exp}). We can write  $\sigma(y,y')$ as follows:
\be 
\sigma(y,y')=\log\left(\frac{1+\sqrt{1-\xi^2}}{\xi} \right) \;\;\;\;\;\; \xi = \frac{2 u u'}{u^2 +u'^2+|x-x'|^2}.
\ee
The integral will be evaluated for finite $\Delta_0$ and later on sent to zero together with the normalization factor of Gamma functions. Now this is a complicated integral so we divide it  depending on the domain of integration. For some constant $r\gg 1$ we divide the domain of integration of each geodesic integral into three sub-domains: $-\infty<\lambda\leq -r$, $-r<\lambda< r$ and $r\leq \lambda< \infty$, and analogously for $\lambda'$. The product of the two integrals then splits into nine pieces out of which four never contain the sub-domain $(-r,r)$.

 Using (\ref{otto}) we can see that, for these four integrals, the geodesic distance rapidly behaves as 
\be\label{appsigma} 
\sigma(y,y')\approx |\lambda| + |\lambda'| \;\;\;\;\;\ \lambda,\lambda' \gg 1.
\ee
 Remembering that the boundary points have been chosen so that $|x_1-x_2|$ and $|x_3-x_4|$ are smaller then the distances between their central points, the geodesic distance between the two RT surfaces then never approaches zero so that the denominator of the integrand (\ref{exp}) never diverges. Then the integrand can be simplified as 
\[
 \frac{e^{-\Delta_0 \sigma(y(\lambda),y'(\lambda'))}}{1-e^{-2\sigma(y(\lambda),y'(\lambda'))}}\xrightarrow[\lambda,\lambda´\gg 1]{} \frac{e^{-\Delta_0 (|\lambda|+|\lambda'|)}}{1-e^{-2(|\lambda|+|\lambda'|)}}\approx e^{-\Delta_0 (|\lambda|+|\lambda'|)}.
\]
In this domain the integral then factorizes as
\[
\int d\lambda\int d\lambda' \frac{e^{-\Delta_0 \sigma(y(\lambda),y'(\lambda'))}}{1-e^{-2\sigma(y(\lambda),y'(\lambda'))}}\xrightarrow[\lambda,\lambda´\gg 1]{} 4\left(\int_{r}^{\infty}d\lambda  e^{-\Delta_0 \lambda}\right)^2.
\]
The integral on the right can be done to give $e^{-r \Delta_0}/\Delta_0\xrightarrow[\Delta_0\rightarrow 0]{}1/\Delta_0$. Analogously $\Gamma(\Delta_0)^2/\Gamma(\Delta_0/2)^4\xrightarrow[\Delta_0\rightarrow 0]{}\Delta_0^2/16$, so overall we obtain the expected result (\ref{exp}) from this term. What about the remaining five integrals containing at least once the region $(-r,r)$? Fortunately we do not need to compute them as the product of Gamma functions in front goes to zero as $\Delta_0^2$ for $\Delta_0\rightarrow 0$ so, if we have not a divergence as $1/\Delta_0^2$ or worst from the integral, the product vanishes. For this reason if one or both of $\lambda,\lambda'$ is restricted to the domain $(-r,r)$ the final result is suppressed and we still obtain (\ref{exp}). 

Before considering in detail $1/c$ corrections let us detour twice. In the first section to study the slightly different problem of computing the "saddle point" contribution to the Renyi entropy for small, but non-zero, value of $\Delta$. In the second section to consider the limit in which one of the two couples of twist and anti twist operators squeezes to zero distance.

\section{Entanglement of purification as saddle point} \label{purification}
 Using the form for the bulk to bulk propagator (\ref{integ})  and noting, once more, that the geodesic distance in our choice of OPE channel is never small, we can approximate it, for $\Delta<<1$, as 
\be \label{simp}
G_{bb}^{\Delta}(y,y')=\frac{e^{-\Delta \sigma(y,y')}}{1-e^{-2 \sigma(y,y')}}\approx e^{-\Delta \sigma(y,y')}.
\ee 
We can also do a second approximation for the integral:
\be
 \int \int e^{-\Delta \sigma}\approx C_{slow} e^{-\Delta \sigma_{min}}+\cdots
\ee
where $C_{slow}$ stands for the integral domain such that the geodesic distance varies sufficiently slow (depending on the required approximation), $C_{slow}=\int \int_{\sigma\approx \sigma_{min}}$ and it does not depend on $\Delta$, and the dots are for the remaining integral whose integrand becomes exponentially smaller. If the propagating operator scaling dimension $\Delta$ is small enough to have $\Delta<\sigma_{min}^{-1} $  then 
\be \label{rem}
S^n(AB)|_{\Delta}\approx -\left(|x_1-x_2|^2\right)^{-\Delta_n}\left(|x_3-x_4|^2\right)^{-\Delta_n}\beta^{-2}_{\Delta} C_{\Delta\; 12}C^{\Delta}_{34}C_{slow} (1-\Delta\sigma_{min}+\cdots )+\dots,
\ee
where we have indicated as $\beta^{-2}_{\Delta}$ the proportionality coefficient inside $W_{\Delta}=\beta^{-2}_{\Delta}\mathcal{W}_{\Delta}$. 

We recall now the recent work \cite{Takayanagi:2017knl} where the authors argue that a quantity called entanglement of purification $E_p(AB)$ \footnote{ for the proper quantum mechanical definition see either \cite{Takayanagi:2017knl} or the discussion in the conclusions } admits an holographic description  as the length of $\Sigma_{min}^{AB}$, which is either the minimal distance connecting the two RT surfaces $\sigma_{min}^{AB}$ when the entanglement wedge is connected, or zero when disconnected:
\be
E_p(AB)=\frac{\Sigma_{min}^{AB}}{4 G_N}.
\ee
 So in the case of connected entanglement wedge (the RT surfaces cross from $\partial A$ to $\partial B$) this is exactly what we have found here as a contribution to $S^n(AB)|_{\Delta}$, as the two geodesics $\gamma_{12}$ and $\gamma_{34}$ are exactly the above mentioned RT  curves: $\sigma_{min}=\sigma_{min}^{AB}=\Sigma_{min}^{AB}$. When instead we are on the other side of the phase transition and the entanglement wedge is disconnected, we can identify $\sigma_{min}$ as $\Sigma_{min}^{(AB)^c}$, and our term inside (\ref{rem}) as proportional to the entanglement of purification of the complement of $A\cup B$ (provided the original state is pure and we do not have a black hole geometry).
 
 The statement would then be that, in computing the two intervals Renyi entropies, part of the total contribution per propagating primary operator is given by a term proportional to the corresponding entanglement of purification, if the scaling dimension is small enough. Then, from the total result for the Renyi entropies, the piece proportional to the entanglement of purification becomes:
 \be \label{rem2}
 E_p(AB) \;(\; \text{or} \; E_p((AB)^c)\;) \propto  \sigma_{min}C_{slow} \cdot \sum_{\Delta<\sigma_{min}}\Delta \beta^{-2}_{\Delta} C_{\Delta\; 12}C^{\Delta}_{34}  .
 \ee
  It is important to point out that a similar result had been already achieved in \cite{Hirai:2018jwy}, although there the description as a saddle point of a gWds was postulated from two assumptions: first that the entanglement of purification can be computed as an expectation value of four non identified operators, and second that these operators have a conformal block expansion whose intermediate channel contains a twist operator. Here we instead followed a reverse path starting from Renyi entropies and identifying certain terms in their dual description as the holographic entanglement of purification.

\section{From four to two}\label{fftt}
To better understand the origin of the relation between $1/c$ quantum bulk corrections to the entanglement entropy and gWds of low scaling operators, from a purely bulk perspective, it is instructive to see how we recover the single interval entanglement entropy from the two intervals case. At the CFT level we achieve this by simply collapsing a twist anti-twist couple into the identity operator: $\braket{\tau\tau^{-1}\tau\tau^{-1}}\rightarrow\braket{\tau\tau^{-1}Id}$. In this holographic setup we will consider the gWd for the identity block in a corresponding limit: one of the two geodesics, say $\gamma_{34}$, collapses to the boundary point where the identity operator is inserted, say $x_0=\frac{x_3+x_4}{2}$.  
So we remain with two bulk to boundary propagators going from $x_1$ and $x_2$ to a generic point along $\gamma_{12}$ and the bulk to bulk propagator going from there to the location of the identity operator, up to a bulk-cutoff distance of $u=\epsilon$; the remaining two bulk to boundary propagators are squeezed together and go from this point where the bulk propagator approaches the identity, $(u=\epsilon,x_0)$, to the actual boundary points  of the collapsing twist and anti-twist operators, $x_0\pm\tilde{\epsilon}$ with $\tilde{\epsilon}$ a boundary cutoff distance. Each of these boundary to bulk propagators will carry an operator with dimension $\Delta_0/2$ ("half of the identity" for each), in the limit $\Delta_0\rightarrow 0$. We also include a normalization $\tilde{\beta}_0$ that should not necessarily agree with the original $\beta^{-2}_0$. Thus:
\be 
S^n(A)|_{\Delta_0}=-\tilde{\beta}_0 C_{12}^0\int_{\gamma_{12}}\hspace{-0.3cm}d\lambda \;G_{b\partial}^{\Delta_n}(y(\lambda),x_1)G_{b\partial}^{\Delta_n}(y(\lambda),x_2)G_{bb}^{\Delta_0}(y(\lambda),\epsilon)G_{b\partial}^{\Delta_0/2}(\epsilon,x_0-\tilde{\epsilon}/2)G_{b\partial}^{\Delta_0/2}(\epsilon,x_0+\tilde{\epsilon}/2)=
\ee 
\[
= \tilde{\beta}_{0}\left(|x_1-x_2|^2 \right)^{-\Delta_n}\left(|\tilde{\epsilon}|^2 \right)^{-\Delta_0 /2} \int_{\gamma_{12}}d\lambda \;G_{bb}^{\Delta_0}(y(\lambda),(x_0,u'=\epsilon)).
\]
The geodesic distance $\sigma$ behaves as (small $\epsilon$):
\[  
\sigma(y(\lambda),(x_0,u'=\epsilon))=\log\left(1+\sqrt{1-\xi}\right)-\log(\xi)\;\;\;\;\;\; \xi = \epsilon \frac{2 u }{u^2 +|x-x_0|^2}
\]
so
\[  
\sigma(y(\lambda),(x_0,u'=\epsilon))\approx -\log(\epsilon)-\log(u(\lambda\rightarrow \pm\infty))
\]
with the usual $\log(u(\lambda))\xrightarrow[\lambda\pm \infty ]{} |\lambda|$ behaviour that, inside $e^{-\Delta_0 \sigma}$ and integrated, behaves as $1/\Delta_0$. So it remains
\[
\left(|x_1-x_2|^2 \right)^{-\Delta_n} \tilde{\epsilon} ^{-\Delta_0}\epsilon^{\Delta_0}\tilde{\beta}_0 \frac{1}{\Delta_0}.
\]
We see then two things. First that, in order to obtain a Renyi entropy matching the RT formula for a single interval, the boundary cutoff should agree with the bulk cutoff, $\epsilon=\tilde{\epsilon}$. This is similar to the usual agreement between the CFT cutoff used in regularizing the entanglement entropy and the holographic cutoff employed in the RT formula. Second that the normalization factor should behave as $\tilde{\beta}_0 \xrightarrow[\Delta_0 \rightarrow 0]{}\Delta_0$ which means something like $\tilde{\beta}_0=4\Gamma(\Delta_0)/\Gamma(\Delta_0/2)^2$ ($\approx$  $\beta_0^{-1}$). We than reduce to the single interval Renyi entropy result:
\be 
\lim_{\Delta_0\to 0}S^n(A)|_{\Delta_0}=\left(|x_1-x_2|^2 \right)^{-\Delta_n}.
\ee

Important here is to notice that in the one interval case the RT formula matches the \emph{exact} CFT one interval computation (beside the $c_n$ constant and UV cutoff); this means that quantum bulk effects in $1/c$, that are non zero for a two intervals entangling region, should be completely suppressed in the limit where one of the two boundary intervals squeezes to zero. Here  what  happens is that, when a couple of twist and anti-twist operators merges into the identity, from a full spectrum of propagating primary operators, we reduce to the identity alone as the only contributing propagating operator, and this term alone reproduces the RT result and thus the total contribution. This is consistent with our previous consideration that quantum bulk effects should be identified (in part) with gWd from higher order conformal blocks. 

\section{Quantum bulk corrections}\label{interpret}
It is time to understand more in depth how holographic corrections to the RT formula emerge from higher order gWds and the backreaction on the geometry.

\subsection{An $O(n-1)$ expansion}\label{expansion}
Let us consider once again (\ref{alt}) including spin to have the full answer at CFT level:
\be 
S(AB)=-\partial_n \left[\sum_{\Delta,l} C_{\Delta,l \; 12}C^{\Delta,l}_{34} W_{\Delta,l}\right]_{n=1}= -\partial_n\left[\sum_{\Delta,l}\beta_{\Delta,l}^{-2} C_{\Delta,l \; 12}C^{\Delta,l}_{34} \mathcal{W}_{\Delta,l}\right]_{n=1},
\ee
where we have included a proportionality factor on the right generalizing the $l=0$ result $\beta_{\Delta,0}^{-2}=\frac{4\Gamma(\Delta)^2}{\Gamma(\frac{\Delta}{2})^4}$. In proximity of $n\approx 1$ the holographic realization of the Renyi entropies entering the above formula is given by the three terms of (\ref{expn1}), one at order $O(1)$ (the identity block in AdS) and two at order $O(n-1)$. Once the analytic continuation for both the $O(n-1)$ terms has been achieved, in principle the complete holographic entanglement entropy can be obtained as:
\be \label{fin}
S(AB)= -\partial_n \Big[\left(|x_1-x_2|^2\right)^{-\Delta_n}\left(|x_3-x_4|^2\right)^{-\Delta_n}\cdot 
\ee
\[
\cdot\sum_{\Delta,l}\beta_{\Delta,l}^{-2} C_{\Delta,l \; 12}C^{\Delta,l}_{34}\int_{\gamma_{12} }\hspace{-0.2cm}d\lambda \int_{\gamma_{34}}\hspace{-0.2cm}  d\lambda'\;G_{b,b}^{\Delta,l}(\lambda,\lambda')|_{O(n-1)}\;+\;\text{backreaction}_{O(n-1)}\Big]_{n=1}.
\]
Once again the term "backreaction$_{O(n-1)}$" symbolizes the computation of the holographic dual to the identity exchange in the backreacted geometry, computed at order $O(n-1)$ before taking the derivative. We will come back to this. The other terms are the standard gWds in AdS, at order $O(n-1)$. Doing the derivative we get (for simplicity we will skip from now on the reminder  $O(n-1)$):
\be
S(AB)= \frac{c}{3}(\log |x_{12}|+\log |x_{34}|)\cdot \left(\sum_{\Delta,l} \dots \right)_{n=1}-
\ee
\[
-\partial_n\Big[\sum_{\Delta,l}\beta_{\Delta,l}^{-2} C_{\Delta,l \; 12}C^{\Delta,l}_{34}|\int_{\gamma_{12} }\hspace{-0.2cm}d\lambda \int_{\gamma_{34}}\hspace{-0.2cm} d\lambda'\;G_{b,b}^{\Delta,l}(\lambda,\lambda')\Big]_{n=1} -\partial_n\left[ \text{backreaction} \right]_{n=1}.
\]
The multiplicative factor in parenthesis in the first line of the above formula has been already computed to give one in section \ref{nc}, by observing that at $n=1$ the twist operators become identity insertions, so that the only OPE term generated is just the identity itself; the corresponding gWd  stripped of the bulk to boundary propagators was then evaluated starting from equation (\ref{exp}). An alternative, easier derivation, comes from the identification of this factor with  $Tr\rho^n|_{n=1}$  stripped of the four bulk to boundary propagators.  As the backreaction term vanishes in the limit $n\rightarrow 1$, and being $Tr \rho=1$, the integral just equals one over the four bulk to boundary propagators at $n=1$:
\[
\left(\sum_{\Delta,l}\dots \right)_{n=1}=\left(|x_1-x_2|^2\right)^{\Delta_{n=1}}\left(|x_3-x_4|^2\right)^{\Delta_{n=1}}=1.
\]

So we have
\be\label{res}
S(AB)= \frac{c}{3}(\log |x_{12}|+\log |x_{34}|)-\partial_n\Big[\sum_{\Delta,l}\beta^{-2}_{\Delta,l} C_{\Delta,l \; 12}C^{\Delta,l}_{34}\int_{\gamma_{12} }\hspace{-0.2cm}d\lambda \int_{\gamma_{34}}\hspace{-0.2cm} d\lambda'\;G_{b,b}^{\Delta,l}(\lambda,\lambda')\Big]_{n=1} -
\ee
\[
-\partial_n\left[ \text{backreaction}(n-1) \right]_{n=1}.
\]
The first term  is again the Ryu-Takayangi formula, the novelty are the two remaining terms that provide 
$1/c$ corrections, being the holographic counterpart of light conformal blocks after insertion of twist operators, and considering back-reaction effects, at order $O(n-1)$. 

Acting with the total $n$ derivative, in the following to avoid confusion indicated as $\frac{d}{d n}$, we get two distinct contributions: on one side derivative terms coming from the dependence of the object by the propagating operator dimension (and spin) when $\Delta=\Delta(n)$, on the other terms that depend on $n$ directly:
\[
\frac{d}{d n}=\frac{\partial}{\partial n}+\partial_n \Delta(n)\frac{\partial}{\partial \Delta(n)}.
\]
A key point now is that the second term in (\ref{res}), when stripped of the OPE structure constants, depends on $n$ only through $\Delta(n)$! This because the geodesic shape is not affected by $n$ (no back reaction for this term), the bulk to bulk propagator do depend only on the scaling dimension (and spin) and the remaining coefficients $\beta_{\Delta}$ do behave in the same way. Obviously the conformal block can have a more general dependence on $n$, but in our holographic description this is absorbed in the  back reaction term. So:
\be \label{der}
-\frac{d}{d n}\left[\sum_{\Delta}\beta^{-2}_{\Delta} C_{\Delta \; 12}C^{\Delta}_{34}\int_{\gamma_{12} } \int_{\gamma_{34}} G_{bb}^{\Delta}\right]_{n=1}=-\sum_{\Delta}\frac{d}{d n}\left(C_{\Delta \; 12}C^{\Delta}_{34}\right)\beta^{-2}_{\Delta}\int_{\gamma_{12} } \int_{\gamma_{34}} G_{bb}^{\Delta}\;|_{n=1}
\ee
\[
-\delta(\Delta(1))C_{\Delta(1) 12}C_{\Delta(1)}^{34}\partial_n\Delta(n)|_{n=1}\partial_{\Delta(n)}\left(\beta^{-2}_{\Delta(n)}\int_{\gamma_{12} } \int_{\gamma_{34}} G_{bb}^{\Delta(n)} \right)_{n=1}.
\]
The delta function in the second term comes from the OPE coefficients at $n=1$ forcing the propagating primary operators with $\Delta=\Delta(n)$ to reduce to the identity in the limit $n\rightarrow 1$: $O_{\Delta(1)}=Id,\; \Delta(1)=0$. We want to show that this last term in fact vanishes. The reason for this is as follows:
\be 
\tau_n(x)\tau_n^{-1}(y)|_{n=1}= Id(y)+\sum_{\tilde{k}} C^{\tilde{\Delta}_k(n)}_{\tau_n \tau_n^{-1}} O_{\tilde{\Delta}_k(n)}(y)|_{n=1},
\ee 
where we named the eventual OPE operators, beside the identity, surviving the $n\rightarrow 1$ limit  as $\tilde{\Delta}_k(n)$. But these should reduce to the identity when $n=1$ so 
\be 
Id(y)+\sum_{\tilde{k}} C^{\tilde{\Delta}_k(n)}_{\tau_n \tau_n^{-1}} O_{\tilde{\Delta}_k(n)}(y)|_{n=1}= Id(y)\left(1+\sum_{\tilde{k}} C^{0}_{Id Id}\right).
\ee 
The term in parenthesis should  go to one, otherwise there would be a discontinuity in the normalization of the two point functions for $n=1$, requiring 
\[
\sum_{\tilde{k}} C^{0}_{Id Id}=0.
\]
But being all the structure constants of definite sign in a unitary theory, each of them should vanish.  So in the limit $n\rightarrow 1$ we indeed find only "one" identity and for all the OPE constants in the last term of (\ref{der}) $C^{\Delta(1)=0}_{ij}|_{n=1}=C^0_{Id Id}=0$. Then (\ref{res}) further simplifies:
\be\label{ipo}
S(AB)= \frac{c}{3}(\log |x_{12}|+\log |x_{34}|)-\sum_{\Delta}\frac{d}{d n}\left(C_{\Delta \; 12}C^{\Delta}_{34}\right)\beta^{-2}_{\Delta}\int_{\gamma_{12} } \int_{\gamma_{34}} G_{bb}^{\Delta}\;|_{n=1}-
\ee
\[
-\partial_n\left[ \text{backreaction}(n-1) \right]_{n=1}
\]

and the key problem to study $1/c$ corrections becomes to compute (apart from the back-reaction part on which we will not dwell in this paper)
\be \label{opeder}
\frac{d}{d n}\left(C_{\Delta \; 12}C^{\Delta}_{34}\right)_{n=1}.
\ee

\subsection{Making contact with the Faulkner, Lewkowycz and Maldacena proposal for quantum corrections}\label{flmquantum}
The Faulkner, Lewkowycz and Maldacena (FLM) proposal \cite{Faulkner:2013ana} for quantum corrections at order $G_N^0$ states that, for a CFT with holographic dual and generic entangling region, indicated as $A$ (either connected or not):
\be \label{flm}
S(A)=S_{RT} +S_{quantum}+O(G_N)
\ee
with
\be\label{flm2}
S_{quantum}=S_{bulk}+\frac{\delta A}{4G_N}+\braket{\Delta S_{Wald-like}}+S_{counter}.
\ee
$S_{RT}$ is the classical RT formula while $S_{bulk}$ computes the \emph{bulk} entanglement entropy between the inside and outside of the RT surface(s) for $A$. We want here to connect this result to our formalism. 

The way (\ref{flm}) was derived is by considering the smooth geometry dual to the CFT replica trick manifold, with compactified Euclidean time circling around the boundary of the entangling region, and computing the bulk partition function in such a background. The classical action is the RT formula \cite{Lewkowycz:2013nqa} while the determinant is:
\be \label{q}
S_{quantum}=\partial_n\left(\log Tr \rho_n^n -n \log Tr \rho_1 \right)_{n=1}
\ee 
with 
\be \label{rho}
\rho_n=P e^{\int_0^{2\pi}d\tau H_n} \;\;\;\;\; \rho_n^n=P e^{\int_0^{2\pi n}d\tau H_n},
\ee
and $H_n$ the bulk Hamiltonian for the Euclidean time circle. The derivative in $n$ then produces two terms, one when acting on the exponent of $\rho_n^n$, that is on the boundary domain of the time integral $2\pi n$, and one when acting on the lower index of $\rho_n$, that is on the Hamiltonian $H_n$ that depends on $n$ as the \emph{smooth} geometry dual to the Renyi entropy depends on $n$  \footnote{for an alternative choice of analytic continuation see the discussion in \cite{Prudenziati:2014tta}}; the first term produces  $S_{bulk}$ while the other amounts to the remaining terms in (\ref{flm2}).

It is both an interesting problem and a check for our result to match the formula for $1/c$ corrections  (\ref{res}) to the above expressions (\ref{q}) and (\ref{rho}), when $d=2$ and $A$ is made of two disconnected intervals. This can be done quite naturally by first isolating terms that depend on $n$ "geometrically" on both sides, that is:
\be \label{dots}
\frac{\delta A}{4G_N}+\braket{\Delta S_{Wald-like}}+S_{counter}=-\partial_n\left[ \text{back reaction}(n-1) \right]_{n=1}
\ee
so that by exclusion
\be \label{bulk}
S_{bulk}=-\partial_n\Big[\sum_{\Delta,l}\beta^{-2}_{\Delta,l} C_{\Delta,l \; 12}C^{\Delta,l}_{34}\int_{\gamma_{12} }\hspace{-0.2cm}d\lambda \int_{\gamma_{34}}\hspace{-0.2cm} d\lambda'\;G_{b,b}^{\Delta,l}(\lambda,\lambda')\Big]_{n=1}.
\ee
In order to understand better equation (\ref{bulk}) we introduce a bulk Renyi entropy $S_{bulk}^n$ such that
\be \label{renbul} 
S_{bulk}=\partial_n S_{bulk}^n|_{n=1}.
\ee 
Then a sufficient condition to verify (\ref{bulk}) is that the Renyi entropies for all $n$ can be expressed as
\be \label{bulkren}
S_{bulk}^ n=-\sum_{\Delta,l}\beta^{-2}_{\Delta,l} C_{\Delta,l \; 12}C^{\Delta,l}_{34}\int_{\gamma_{12} }\hspace{-0.2cm}d\lambda \int_{\gamma_{34}}\hspace{-0.2cm} d\lambda'\;G_{b,b}^{\Delta,l}(\lambda,\lambda').
\ee
This last formula (\ref{bulkren}) will be the focus of the present section. 

 When we want to compute Renyi entropies for a generic two dimensional quantum field theory  we know we can express the problem as correlators of twist and anti-twist operators at the boundary of the entangling surface. In three dimensions this boundary is a line, so we would need twist lines. A formalism for twist line operators has not yet been fully developed (to the author's knowledge) so we will proceed with caution.

The property that a twist line $\tau_n(\gamma)$ should have is that fields circling around  either increase or decrease their replica index. To determine the direction of rotation we associate an orientation and fix conventions such that a clock-wise and counter-clockwise rotations respectively decrease and increase the replica index, $i\rightarrow i-1$ and $i\rightarrow i+1$, ($n+1=1$). If such a twist line is embedded in the bulk and it intersects (not tangentially) any two dimensional slice $H_2$ at some set of points, $\gamma\cap H_2=\{x_1, \cdots x_s\}$, fields living on $H_2$ will see $\{x_1, \cdots x_s\}$ as twist or anti-twist \emph{operator} locations, depending on the local orientation of $\tau_n(\gamma)$ \footnote{ A tangential point correspond to the identity operator}. If we choose $H_2$ to be the AdS conformal boundary, then we have constructed usual twist and anti-twist operators in a two dimensional CFT, out of one (or multiple) twist line(s) living in the holographic dual theory and intersecting the conformal boundary at the twist and anti-twist operator positions.

To compute the Renyi entropies $S^n_{bulk}$ across the RT surfaces $\gamma_{12},\gamma_{34}$  for the two intervals \emph{boundary} entangling region $A\cup B$, we need to compute the two twist lines correlator for the \emph{bulk} Renyi entropies:
\be \label{twist}
S_{bulk}^n=-\braket{\tau_n(\gamma_{12}) \tau_n(\gamma_{34})}=-\braket{\int_{\gamma_{12}}^{1\rightarrow 2}\tau_n(s)ds \int_{\gamma_{34}}^{3\rightarrow 4}\tau_n(s')ds'},
\ee
where the orientation is written on the top of the integral. This orientation is fixed, modulo a global flip, such that when the two twist lines are made to coincide they should annihilate each other: here this would be achieved by sending $x_1\rightarrow x_4$ and $x_2\rightarrow x_3$, that means either a zero or space filling entangling region, which in turn implies vanishing entanglement entropy for the CFT in a pure state. As a consequence a closed path linking just  $\gamma_{12}$ or $\gamma_{34}$ raises or lowers the replica index of the dragged bulk field, but if the path does not link any of the RT surfaces or links both, the monodromy is trivial.

Following our identification (\ref{bulk}) we would like to propose the following result:
\be \label{tyu}
\braket{\int_{\gamma_{12}}^{1\rightarrow 2}\tau_n(s)ds \int_{\gamma_{34}}^{3\rightarrow 4}\tau_n(s')ds'}=\sum_{\Delta,l}\beta_{\Delta,l}^{-2} C_{\Delta,l \; 12}C^{\Delta,l}_{34}\int_{\gamma_{12} }\hspace{-0.2cm}d\lambda \int_{\gamma_{34}}\hspace{-0.2cm} d\lambda'\;G_{b,b}^{\Delta,l}(\lambda,\lambda').
\ee
This formula states that, to compute the correlator $\braket{\tau_{\gamma_{12}} \tau_{\gamma_{34}}}$, the connected tree-level bulk diagrams appearing contains a single bulk to bulk propagator in between generic points of $\tau_{\gamma_{12}}$ and  $\tau_{\gamma_{34}}$  \footnote{it is perhaps important to point out that it has been conjectured in \cite{Hijano:2015zsa} that not only tree level bulk diagrams but also loop diagrams can be decomposed in gWd, that are always tree level by construction. This seems to suggest that the validity of (\ref{tyu}) goes beyond tree-level. }. 
 
 The important point here, in order to understand (\ref{tyu}), is to figure out what fields do interact with the two twist lines and with what interaction. We have seen that we can interpret a couple of twist and anti-twist operators in the boundary CFT as the intersection of the boundary with a single bulk twist line. Moreover, when the twist line is along the RT surface connecting these two points, the corresponding replicated bulk geometry is dual to the boundary state produced by the path integral on the replica trick space; this has been shown in \cite{Lewkowycz:2013nqa}. In this sense the interaction between two \emph{couples} of twist and anti-twist fields  in the CFT, as we found in our OPE expansion of conformal blocks, should then contain intermediate propagating operators that are exactly the duals of the bulk fields for the corresponding bulk twist lines correlator. This is the simple explanation of (\ref{tyu}). Then we can immediately state the Feynman rules for such an interaction: when we insert a propagator $G_{b,b}^{\Delta,l}$ in between the two twist lines each vertex is written as $C_{\Delta,l}/\beta_{\Delta,l}$, where $O_{\Delta,l}$ and $C_{\Delta,l}$ are  the same primary operators and OPE coefficients appearing in the conformal block expansion of two couples of twist and anti-twist operators. 

Before testing our proposal it is important to discuss the issue of renormalization and its effects on the interpretation of (\ref{dots}) and (\ref{bulk})-(\ref{bulkren})-(\ref{tyu}) \footnote{I thank the referee for mentioning this important point.}. When computing the bulk entanglement entropy this gets renormalized by the insertion of counterterms included here in (\ref{dots}), which come from the usual renormalization procedure of the gravitational action. These counterterms makes the bulk entanglement entropy finite absorbing the divergent terms into the renormalized couplings, among these the Newton constant $G_N$ \cite{Solodukhin:2011gn}, \cite{Susskind:1994sm} and \cite{Jacobson:1994iw}. Through the relationship of $G_N$ with the boundary central charge $c$, $c=3R/2G_N$, this means that the central charge as well goes through renormalization. Thus, for instance, the tree level boundary result for a single interval $S(l)=\frac{c}{3}\log(\frac{c}{\epsilon})$ transforms via the induced renormalization of $c$, once quantum corrections are included. For this reason the identifications (\ref{dots}) and (\ref{bulk})-(\ref{bulkren})-(\ref{tyu}) should always be  computed after the renormalization has been done. In the one interval case, once more, this means that the bulk quantum corrections are really the difference, at order $G_N^0$, with respect to  the tree level result $S(l)=\frac{c}{3}\log(\frac{c}{\epsilon})$ with the central charge $c$ renormalized.

With this in mind let us face an interesting problem to test our formalism, to compute the bulk Renyi entropies (understood as above) for the single twist line along the RT surface corresponding to a \emph{single} interval entangling region $A$. We know that this should ultimately gives zero because the single interval \emph{boundary} entanglement entropy $S(A)$ in the CFT vacuum state already matches the classical RT formula, so no quantum bulk corrections and in particular no bulk entanglement should be found in this case \footnote{ this statement is of course no longer valid for the CFT in a generic state, as for example in an exited state, see \cite{Belin:2018juv}, or a thermal state.}. Alternatively, following \cite{Barrella:2013wja}, we can study the quantum bulk corrections as a functional determinant around the bulk geometry $AdS/\Gamma$ dual to the boundary Riemann surface replica manifold $\mathbb{C}/\Gamma$. In this language quantum corrections are zero  because for a single interval the uniformizing map maps the replica $n$-sheethed manifold to the full complex plane, without doing any quotient by $\Gamma$, so the final genus is zero. So the formula for quantum corrections (36) of \cite{Barrella:2013wja} is also zero.
Thus using the language of gWds the result that we want to reproduce is, being $\gamma$ any single connected RT curve inside AdS:
\be \label{quest}
\braket{\int_{\gamma}\tau_n(s)ds}=0.
\ee
Due to the complicated form of the bulk to bulk propagator for fields dual to operators with spin, this is quite complicated to check directly, unless we apply a few tricks. First we note that, following our idea, (\ref{quest}) should be computed by a Feynman graph with a bulk to bulk propagators connecting any two points of the twist line along $\gamma$.  Given the Feynman rules derived above this is:
\be \label{s1}
\braket{\int_{\gamma}\tau_n(s)ds}=\sum_{\Delta,l}\beta_{\Delta,l}^{-2} C_{\Delta,l }^ 2\int_{\gamma }d\lambda \int_{\gamma}d\lambda'\;G_{b,b}^{\Delta,l}(\lambda,\lambda').
\ee
The second step is to express this double integral as a limit of a usual gWd with stripped bulk to boundary propagators, where two couples of  boundary points approach each other. For example if we pick $\gamma$ to be $\gamma_{13}$, then sending $x_1\rightarrow x_2$ and  $x_3\rightarrow x_4$ the geodesics  $\gamma_{24}\rightarrow \gamma_{13}=\gamma$ \footnote{from the point of view of a twist line correlator the two twist lines approach each other in this limit and, due to opposite orientation, they should annihilate each other, as already discussed. This is consistent with our claim (\ref{quest}) and with the result we will find, however for the moment we are just considering the two integrals in (\ref{s1}) where the result is not yet clear.}. Then we can write (\ref{s1}) as
\be \label{primo}
\braket{\int_{\gamma}\tau_n(s)ds}=\lim_{x_{1/3}\rightarrow x_{2/4}} \sum_{\Delta,l}\frac{C_{\Delta,l}^2}{\beta_{\Delta,l}^2} \int_{\gamma_{13} }\hspace{-0.2cm}d\lambda \int_{\gamma_{24}}\hspace{-0.2cm}d\lambda'\;G_{b,b}^{\Delta,l}(\lambda,\lambda')=\lim_{x_{1/3}\rightarrow x_{2/4}} \sum_{\Delta,l}C_{\Delta,l}^2 W_{\Delta,l}(x_i) |x_{13}|^{2\Delta_n}|x_{24}|^{2\Delta_n},
\ee 
where in the last equality we made use of (\ref{bp}). Note that with $x_1\rightarrow x_2$ and  $x_3\rightarrow x_4$ the shortest geodesics would be $\gamma_{12}$ and $\gamma_{34}$, but we instead keep $\gamma_{13}$ and $\gamma_{24}$ as the total result should not depend on the channel of OPE. The third trick is to compare the above result with the conformal blocks expansion in the usual channel; starting from
\be 
\sum_{\Delta,l}C_{\Delta,l}^2 W_{\Delta,l}=\sum_{\Delta,l}C_{\Delta,l}^2 \frac{ G_{\Delta,l}(u,v)}{ |x_{12}|^{2\Delta_n}|x_{34}|^{2\Delta_n}},
\ee 
we get:
\be \label{secondo}
\braket{\int_{\gamma}\tau_n(s)ds}=\lim_{x_{1/3}\rightarrow x_{2/4}}\sum_{\Delta,l}C_{\Delta,l}^2  G_{\Delta,l}(u,v)u^{-2\Delta_n}.
\ee
We can now use the exact results for conformal blocks in $d=2$, with or without spin, for external scalar operators \cite{Dolan:2003hv}, \[G_{\Delta,l}(z,\bar{z})=|z|^{\Delta-l}\left[z^l\cdot\prescript{}{2}F_1\left(\frac{\Delta+l}{2},\frac{\Delta+l}{2};\Delta+l,z\right)\prescript{}{2}F_1\left(\frac{\Delta-l}{2},\frac{\Delta-l}{2};\Delta-l,\bar{z}\right)+c.c.\right], \] having defined $z,\bar{z}$ from the equations \[ u=z\bar{z}, \;\;\; v=(1-z)(1-\bar{z}) \]  that can be explicitly solved as \[z=\sqrt{u}\exp(i \;\text{ArcCos}\left(\frac{1-v+u}{2\sqrt{u}}\right)),\;\;\; \bar{z}=\sqrt{u}\exp(-i \;\text{ArcCos}\left(\frac{1-v+u}{2\sqrt{u}}\right)). \] 
The limit $x_1\rightarrow x_2$ and  $x_3\rightarrow x_4$ translates first into $u\rightarrow 0$ and $v\rightarrow 1$ and then into $z\rightarrow i \sqrt{u}$ and $\bar{z}\rightarrow -i \sqrt{u}$. Then in this limit
\be 
G_{\Delta,l}(z,\bar{z})\rightarrow 2 u^{\Delta/2}\text{Cos}\left(\frac{l\pi}{2}\right),
\ee
which means
\be 
\braket{\int_{\gamma}\tau_n(s)ds}\rightarrow \sum_{\Delta,l}C_{\Delta,l}^2 u^{\Delta/2-2\Delta_n}\text{Cos}\left(\frac{l\pi}{2}\right).
\ee
Each of these terms is separately zero in the $u\rightarrow 0$ limit whenever $\Delta>4\Delta_n$, which happens always in the limit $n\rightarrow 1$. Also the derivative at $n=1$ produces  
\be 
\partial_n\braket{\int_{\gamma}\tau(s)ds}_{n=1}\rightarrow -\sum_{\Delta,l}C_{\Delta,l}^2 \frac{c}{3}\log(u) u^{\Delta/2}\text{Cos}\left(\frac{l\pi}{2}\right),
\ee
which is zero when $u\rightarrow 0$. This proves also in the gWd formalism that the bulk entanglement entropy for a single region is zero, thus supporting our claim (\ref{bulk}) and (\ref{tyu}) \footnote{ It is important to note that the expansion in the OPE is different between (\ref{primo}) and (\ref{secondo}), so that the single terms in the sum of (\ref{primo}) do not necessarily vanish.}.

\subsection{Twist operators OPE at lowest order (beside the identity)}\label{ccts}
We borrow results from the literature to compute  the  analytic continuation in $n$ for the OPE coefficients (\ref{opeder}) entering (\ref{ipo}), for the lowest primary of scaling $\Delta$. This computation is a nice example and also contains useful results for the following section, where the mutual information is considered.

In \cite{Calabrese:2009ez} and \cite{Calabrese:2010he} the authors consider the replica trick partition function for a given set of intervals, and replace one of these intervals with a "generalized" OPE expansion given by a sum of products of operators belonging to different sheets in the replica trick manifold. Translated in the twist operator formalism the general OPE between twist and anti-twist operators is written as \footnote{the expansion point $z$ is some mid point anywhere inside the interval $x-y$.}:
\be \label{genope}
\tau_n(x)\tau_n^{-1}(y)=\sum_{\{k_j\}}C_{\{k_j\}}\prod_{j=1}^n O_{k_j}(z)
\ee
where $O_{k_j}$ are a complete set of operators for the $j$th-replica theory (that was the theory on  sheet $j$), $j=1\dots n$. This is a generalization of the usual OPE and it is there used to derive a formula for  (minus) the Renyi entropy $S^n(AB)$ of equation (\ref{ren2}):
\be \label{cct}
Tr \rho_{AB}^n = c_n^2 \left(x_{12}^2 x_{34}^2 \right)^{-\frac{c}{12}(n-\frac{1}{n})}\sum_{\{k_j\}}\left(\frac{x_{12}^2 x_{34}^2}{ r^2} \right)^{\sum_j h_{k_j}+\bar{h}_{k_j}}d^ 2_{\{k_j\}}.
\ee
The coefficients $d_{\{k_j\}}$ can be computed for primary operators to be
\be \label{dk}
d_{\{k_j\}} = n^{-\sum_j h_{k_j}+\bar{h}_{k_j}}\braket{\prod_{j=1}^n O_{k_j}(e^{2\pi i j /n})}|_{\mathbb{C}}.
\ee
Notation is as follows: $c_n$ is the coefficient that expresses the proportionality between the Renyi entropy and the twist field computation, it should obey $c_1=1$ but it is otherwise undetermined \footnote{see footnote \ref{foot}}. The distance $r$ is between the center of the two intervals $A$ and $B$, so $r=\frac{1}{2}|x_1+x_2-x_3-x_4|$ and $h,\bar{h}$ are left and right weights ($\Delta = h + \bar{h}$ and $h-\bar{h}=l$). The expectation value is on the uniformized plane  $\mathbb{C}$ for a \emph{single} interval, that is the space where each sheet is mapped to a wedge of angle $2\pi /n$ and Real infinity on the j-th sheet goes to $e^{2\pi i j/n}$.

How do we transform expression (\ref{cct}) in the conformal block language? The conformal blocks join all the contributions from a primary operator and its descendants, so that writing $\sum_{\{kj \}}=\sum_{\{p_j\}}\sum_{\{k_{p_j} \} }$ with $p_j$ a primary on the sheet $j$ and $\{k_{p_j} \}$ its descendants we have: 
\be \label{cct2}
Tr \rho_{AB}^n = c_n^2 \left(x_{12}^2 x_{34}^2 \right)^{-\frac{c}{12}(n-\frac{1}{n})}\sum_{\{p_j\}}\left(x_{12}^2 x_{34}^2 \right)^{\sum_j h_{p_j}+\bar{h}_{p_j}}\cdot 
\ee
\[
\cdot \left(r^2  \right)^{-\sum_j h_{p_j}+\bar{h}_{p_j}}d^ 2_{\{p_j\}}\sum_{\{k_{p_j}\}}\xi^{p_j\{k_{p_j} \}}_{12}\bar{\xi}^{p_j\{\bar{k}_{p_j} \}}_{12}\left(x_{12}^2 x_{34}^2 \right)^{K_{p_j}+\bar{K}_{p_j}}\left(r^2  \right)^{-K_{p_j}-\bar{K}_{p_j}}t_{\{k_{p_j} \}}
\]
with $K_{p_j}=\sum_{i}k^i_{p_j}$ the additional dimension of the descendant state $\{k_j\}$ written as $L_{-k^1_{p_j}}\dots L_{-k^n_{p_j}}\ket{p_j}$, $d^ 2_{\{k_j\}}=d^ 2_{\{p_j\}}\xi^{p_j\{k_{p_j} \}}_{12}\bar{\xi}^{p_j\{\bar{k}_{p_j} \}}_{12}$ and $t_{\{k_{p_j} \}}$ the additional normalization of the two point function for descendants. Then following notation as in (\ref{uno}), for a primary field $ \{p_j\}$
\be \label{opecoeff}
C^{\{p_j\}}_{ij}= c_n d_{\{p_j\}}
\ee
while the conformal  block $W_{\{p_j\}}(x_i)$ contains all the rest \footnote{It is probably important to point out that the generalized OPE coefficients (\ref{genope}) cannot in general be interpreted as an OPE of the original theory, even for $n=1$.}. 

When we restrict to the contribution in (\ref{genope}) from a product of only two operators, $\{k_j \}=(k_{j_1},k_{j_2})$ with $k_{j_1}=k_{j_2}=k$ to be non zero, an expression for the analytic continuation of the Renyi entropies is available in \cite{Calabrese:2010he}. This is the leading contribution outside the identity block when considering the lowest $h_k$ in the spectrum beside the identity. The expression they found is:
\be \label{ccton}
d_{2,k}(n)^2=\frac{s_2(n)}{(n^2)^{2h_k+2\bar{h}_k}}
\ee
with 
\be 
s_2(n)=\frac{n-1}{2}g(0)+\sum_{m=1}^{\infty}(n g(nm)-g(m))
\ee
and
\be 
g(m)=\frac{4^{\alpha}}{\pi^2}\text{sin}(\pi\alpha)\Gamma(1-2\alpha)\text{sin}(\pi(\alpha-m))\Gamma(\alpha-m)\Gamma(\alpha+m),
\ee
having written $\alpha=2(h_k+h_{\bar{k}})$. Because of equation (\ref{ipo}) this is all we need for the analytic continuation of the corresponding gWd in (\ref{res}). Its $n$ derivative in $n=1$, multiplied by the additional doubly integrated bulk to bulk propagator factor in (\ref{ipo}), gives us the very first correction to the entanglement entropy. 

\subsection{Mutual information}\label{mutual}
The mutual information $I(A,B)=S(A)+S(B)-S(AB)$ is an interesting quantity to consider when looking to quantum corrections to the RT formula. The reason is that, for two well separated intervals, the classical contribution and the local part of the quantum bulk corrections (\ref{dots}) cancel, leaving only the non local bulk entanglement piece (\ref{bulk}):
\be \label{qui}
I(A,B)=S(A)_{bulk}+S(B)_{bulk}-S(AB)_{bulk}=-S(AB)_{bulk}
\ee
where in the second equality we used the result of the previous section for a single interval bulk entanglement entropy. By using (\ref{bulk}) together with (\ref{bp}), (\ref{ipo}), (\ref{cct2}) and (\ref{opecoeff})  we find 
\[ 
I(A,B)\approx \sum_{\Delta}\frac{d}{d n}\left(C_{\Delta(n) \; 12}C^{\Delta(n)}_{34}\right)_{n=1}r^{-2\Delta},
\]
which is the form proposed in the literature, see for example \cite{Agon:2015ftl} and \cite{Faulkner:2013ana}. This is then an additional independent check that the identification  (\ref{bulk}) is indeed meaningful.

\section{Conclusions and future work}\label{cocnclusions}
In this paper we have exploited the  representation of conformal blocks as geodesic Witten diagrams in order to study how the holographic description of the two intervals entanglement entropy emerges. We have seen that the Ryu-Takayanagi formula is reproduced by the four bulk to boundary propagators, which is the full answer when considering the Witten diagram dual to the identity block, while quantum corrections come from stripped higher order diagrams and back-reaction on the geometry from the twist operators. We have matched these terms with the result from \cite{Faulkner:2013ana} and in particular identified the bulk entanglement entropy across the RT surfaces as a sum over the same Witten diagrams entering the conformal block expansion of the four twist operator correlators, but stripped of the bulk to boundary propagators. This has also been derived by using a novel twist line formalism in the bulk and interpreting the Witten diagrams as a point like interaction between the two twist lines. We have also applied the formalism to the single interval problem by showing how quantum corrections vanish in this case. Additional discussions on the analytic continuation of the diagrams and in particular of the OPE coefficients have been included, borrowing literature results from \cite{Calabrese:2009ez}-\cite{Calabrese:2010he}, that have later on been used to study the mutual information, matching the expected form previously proposed in the literature. Finally an interesting by-product result is the unexpected connection between an approximated form of the light gWd correction to the Renyi entropies and the proposed holographic description for the entanglement of purification.

Different directions for future work exist. Among them the most promising ones are listed:

\subsubsection*{Quantum mechanical interpretation of the relation between entanglement entropy and entanglement of purification}
The entanglement of purification $E_p(AB)$ has a quantum mechanical definition as the minimal entanglement entropy
\be 
E_p(AB)= \text{min} \;S(A\tilde{A})
\ee
such that
\be 
 \rho(AB)=\text{Tr}_{\tilde{A}\tilde{B}}\left(\ket{\psi}\bra{\psi}\right),
\ee
where $\rho(A\tilde{A})=\text{Tr}_{B\tilde{B}}\left(\ket{\psi}\bra{\psi}\right)$, and the minimization has been done over any pure state $\ket{\psi}\in \mathcal{H}_{A\tilde{A}}\otimes \mathcal{H}_{B\tilde{B}}$. Its holographic interpretation has been proposed in \cite{Takayanagi:2017knl}. Here we saw that, based exclusively on its bulk description, we could obtain a quantity proportional to  $E_p(AB)$ as a "saddle point" approximation of any term in the Renyi entropies $S^n(AB)$ expansion, provided a sufficiently low scaling dimension of the propagating operator. What is the quantum mechanical interpretation of the above statement is an intriguing question who already received a partial answer in \cite{Tamaoka:2018ned}; there a connection with a quantity $S_o(AB)$ called odd entanglement entropy was made, where the latter is essentially entanglement entropy for a partially transposed reduced density matrix and considering analytic continuation from odd integers $n$. Then the result from \cite{Tamaoka:2018ned}:
\be 
E_p(AB)=S_o(AB)-S(AB)
\ee 
is quite reminiscent of (\ref{rem})-(\ref{rem2}) after derivative in $n$.

It would be interesting to further develop the arguments of \cite{Tamaoka:2018ned} in an attempt to derive the present relation. Alternatively a different path could be to study in depths the MERA tensor network realization of the reduced density matrix $\rho_{AB}$, that also very likely provided the original idea for \cite{Takayanagi:2017knl}, and from there to try to infer the connection.

\subsubsection*{Multi interval entanglement entropy}
An interesting and somehow obvious generalization of this work would be to consider multi-interval entanglement entropy by using gWds with additional legs. Unfortunately  such formalism has not yet been developed and it looks that the decomposition of a higher legs Witten diagram into gWds is quite non-trivial. Beside the original discussion in \cite{Hijano:2015zsa}, recent work in this direction is contained in, among the others, \cite{Alkalaev:2016rjl}, \cite{Alkalaev:2015fbw}, \cite{Banerjee:2016qca} and \cite{Parikh:2019ygo}.

\subsubsection*{Quantum corrections to the holographic entropy cone}
In \cite{Bao:2015bfa} a surprisingly in depth description of the entropy cone for holographic states (compared with the few information we have for generic states) was given, together with an algorithm providing a sufficient condition for any putative entropy inequality to be correct. The paper based its analysis on the classical geometric description of the entanglement entropies translated into a certain graph construction. Indeed it would be very valuable to be able to improve the classical analysis including quantum bulk contributions, perhaps in the form of certain graph corrections based on the Feynman-like rules we described.

\subsubsection*{Entanglement entropy across black hole horizon}
The Renyi entropies across the event horizon for a three dimensional BTZ black hole can, in principle, be computed by considering a twist line correlator and the gWd formalism developed here, only evaluated in a different background. This may reproduce the standard Bekenstein-Hawking black hole entropy at leading order, if the latter is identified with the black hole entanglement entropy, and for higher propagating scaling dimension and spin the  leading corrections for entanglement across the event horizon. To compare these computations with literature results, mainly from direct application of the replica trick to the gravitational partition function (see \cite{Solodukhin:2011gn} for a review), is an interesting problem. In particular I would like to pursue the possible connection between these corrections and \cite{Prudenziati:2018jcf}.

\subsubsection*{A bulk parametric expansion}
The CFT conformal blocks accept a small $x$ expansion when the four operators have been placed at $0,x,1,\infty$ by conformal transformations. Analogously we can look for some small parameter expansion for the corresponding integrals in the bulk. 

With this in mind we consider the case where the distance between the central points of the two intervals, linked by the geodesics $\gamma_{12}, \gamma_{34}$, is bigger then the size of the two intervals. When this happens the parameters  $\gamma(\lambda,\lambda')\equiv \frac{u(\lambda)}{|x(\lambda)-x'(\lambda')|}$ and  $\gamma'(\lambda,\lambda')\equiv \frac{u'(\lambda')}{|x(\lambda)-x'(\lambda')|}$ are smaller then one (along the solutions (\ref{otto})) and we can expand in them. The nice feature is that the quantity $\xi$ on which $\sigma$ depends has a nice simple form when expressed in $\gamma,\gamma'$:
\be 
\xi=\frac{2 \gamma \gamma'}{1 + \gamma^2 + \gamma'^2}.
\ee
We can then  consider the integrand (\ref{integ}) as a function of $\gamma$ and $\gamma'$ and eventually  expand at arbitrary order:
\be\label{inap}
\left(\frac{1}{\gamma \gamma'} \right)^{-\Delta}\big(1-\Delta(\gamma^2+\gamma'^2)+\frac{1}{2}\Delta(1+\Delta)(\gamma^4+\gamma'^4)+(1+\Delta)^2\gamma^2\gamma'^2 - 
\ee
\[
-(2+4\Delta+5\Delta^2/2+\Delta^3/2)(\gamma^2\gamma'^4+\gamma'^2\gamma^4)-\frac{1}{6}\Delta(1+\Delta)(2+\Delta)(\gamma^6+\gamma'^6)+\dots O((\gamma+\gamma')^8) \big).
\]
Note that this expansion goes together with powers of $\Delta$, so it makes sense only for $\Delta<1$. This can be either integrated in $\lambda$, $\lambda'$ at any order numerically (but numerically you can even integrate the exact expression) or you can choose to express the integrals in the variables $\gamma$, $\gamma'$ itself. This change of variables brings the Jacobian:
\be
|\det \left(\frac{\gamma}{\lambda}\right)^{-1}|=\left|\gamma \gamma' \tanh(\lambda)\tanh(\lambda')(1+\frac{\gamma}{\sinh(\lambda)}-\frac{\gamma'}{\sinh(\lambda')})\right|^{-1}
\ee
so that we need to express the hyperbolic functions as functions of $\gamma$; this is not trivial in the desired form and can be done easily only by fixing either $\lambda$ or $\lambda'$ to zero. In this case   however it is not clear how to use them, as the integrals should be computed in $\gamma$ and $\gamma'$ unconstrained. It is an interesting future problem to work out in detail such an expansion, or alternative possibilities.

\section*{Acknowledgments}
I would like to thank Jacopo Viti and Tadashi Takayanagi for discussion and for having pointed out an interesting reference that I had missed.

This work has been done under financial support from the Brazilian ministries MCTI and MEC.

\end{document}